\documentclass[aps,prd,twocolumn, notitlepage,superscriptaddress,10pt,
amsmath,amssymb,nofootinbib]{revtex4-1}

\usepackage{color}
\usepackage{graphicx}

\newcommand{\fett}[1]{\boldsymbol{#1}}

\newcommand{\dd}{{\rm{d}}}

\newcommand{\be}{\begin{equation}}
\newcommand{\ee}{\end{equation}}

\newcommand{\ws}{\textcolor{white}{!}}
\newcommand{\nabq}{\fett{\nabla}_{\fett{q}}}
\newcommand{\nabx}{\fett{\nabla}_{\fett{x}}}

\definecolor{darkred}{rgb}{0.5,0,0}
\definecolor{darkgreen}{rgb}{0,0.5,0}
\definecolor{darkblue}{rgb}{0,0,0.5}

\usepackage{hyperref}
\hypersetup{colorlinks,
linkcolor=darkblue,
filecolor=darkgreen,
urlcolor=darkred,
citecolor=darkblue }

\begin{document}

\preprint{\vbox{\hbox{\hfill TTK-13-12, TUM-888-13}}}

\title{Initial conditions for cold dark matter particles and General Relativity}

\date{\today}

\author{Cornelius Rampf}
\email{c.rampf@unsw.edu.au}
\affiliation{Institut f\"ur Theoretische Teilchenphysik und Kosmologie, RWTH Aachen, D--52056 Aachen, Germany}
\affiliation{School of Physics, University of New South Wales, Sydney, NSW 2052, Australia}

\author{Gerasimos Rigopoulos}
\email{g.rigopoulos@tum.de}
\affiliation{Physik Department T70, Technische Universit\"{a}t M\"{u}nchen, D--85748 Garching, Germany}

\begin{abstract}
We describe the irrotational dust component of the Universe in terms of a relativistic gradient expansion and transform the resulting  synchronous metric to a Newtonian coordinate system. The two metrics are connected via a spacelike displacement field and
a timelike perturbation, providing a relativistic generalization of the transformation from Lagrangian to Eulerian coordinates.
The relativistic part of the displacement field generates
already at initial time a nonlocal density perturbation at second order.
This is a purely relativistic effect since it originates from space-time mixing.
We give two options, the passive and the active approach, on how to include the relativistic corrections for example
in $N$-body simulations.
In the passive approach we treat the corrections as a non-Gaussian modification of the initial Gaussian  field
(primordial non-Gaussianity could be incorporated as well). 
The induced non-Gaussianity depends on scale and the redshift at which initial conditions are set, with $f_{\rm NL} \sim {\rm few}$ for small enough scales and redshifts. In the active approach we show how to use the relativistic trajectory to obtain the initial displacement
and velocity of particles for $N$-body simulations without modifying the initial Gaussian field.
\end{abstract}

\maketitle

\section{Introduction}

Linearized cosmological perturbation theory is a key technique for studying the nature of cosmological inhomogeneities \cite{Mukhanov:1990me,Ma:1995ey}.
Its extension to second order has been applied to: (1) inflation, together with the subsequent reheating \cite{Maldacena:2002vr,Acquaviva:2002ud,Bartolo:2004if}, (2) to the coupled
set of Einstein-Boltzmann
equations of the primordial baryon-photon fluid \cite{Bartolo:2006fj,Beneke:2010eg,Pettinari2013}, (3) to secondary effects after decoupling of the photons  \cite{Hu:1993tc,DeTroia:2003tq},
and (4) to late-time evolution of gravitational clustering \cite{Bernardeau:2001qr,Matarrese:1995sb}.
It is crucial to go to second order to understand nonlinear aspects of the underlying physics, and to disentangle the various sources of non-Gaussianities.

Here we report a somewhat different source of non-Gaussianity, relevant for the initial conditions of the cold darm matter (CDM) cosmic component, which cannot be directly embedded into one of the above groups (1)-(4). Instead,  non-Gaussianity arises because of the nonlinear coordinate transformation from a synchronous metric akin to Lagrangian coordinates to a Newtonian-like coordinate
system, used for example in setting up $N$-body simulations. In Ref.~\cite{Rampf:2012pu} it has been recently shown that such a coordinate transformation connects both metrics with a space-like displacement field
and a timelike perturbation. The spacelike displacement field is strongly related to
the one in the Newtonian Lagrangian perturbation theory (LPT), and the timelike perturbation can be interpreted as the 3-velocity potential of the
displacement field. This correspondence encourages us to interpret the synchronous coordinate system to be Lagrangian, attached to the CDM particles,
whereas the observer's "Eulerian'' position is
at rest in the Newtonian frame. Indeed, the metric perturbations in the Newtonian frame resolve to the Newtonian cosmological potential in the appropriate limit.

Relativistic corrections in the displacement field should not influence the gravitational late-time evolution of the particle trajectories much.
As we shall demonstrate however, the relativistic coordinate transformation
leads to non-Gaussian contributions in the density perturbations, 
which relate to the 3-velocity potential of the displacement field. These corrections matter already at the initial time and can be interpreted as a non-Gaussian modification of
an otherwise initial Gaussian field.

This \emph{paper} is organized as follows. In Sec.~\ref{sec:grad} we first present the gradient expansion metric in the synchronous frame for irrotational and pressureless cold dark matter particles in a $\Lambda$CDM universe, where $\Lambda$ denotes a cosmological constant.
Then, we show
how the residual freedom inherent in the gradient metric can be fixed, clarify its physical interpretation and separate the sources/origin of non-Gaussianities in the latter coordinate transformation, see Sec.~\ref{sec:Newt}.
In Sec.~\ref{sec:ING} we interpret the relativistic corrections as the above mentioned non-Gaussian modification of the initial Gaussian field ("the passive approach'') and approximate them in terms of a bispectrum component; the particles are then displaced according to Newtonian LPT.
In Sec.~\ref{sec:opB} we describe the active approach where the initial Gaussian field is unaltered but the particles
are displaced according to the relativistic trajectory.
Then, in Sec.~\ref{sec:deltaN} we calculate the density contrast
in the Newtonian coordinate system.
Finally, we relate the density contrast in the
Newtonian gauge to the one in $N$-body simulations (Sec.~\ref{sec:which}),
and we conclude in Sec.~\ref{sec:con}.

\ws

\section{The gradient expansion metric}\label{sec:grad}

We use the gradient expansion technique to solve the Einstein equations
\cite{Lifshitz:1963ps,Tomita:1975kj,Tanaka:2006zp,Stewart:1994wq,Enqvist:2011aa,Rigopoulos:2012xj}, although related relativistic approximation schemes lead to similar results \cite{Matarrese:1994wa,Russ:1995eu,Bartolo:2010rw}.
We use a comoving/synchronous line element
\be
  \dd s^2 = - \dd t^2 + \gamma_{ij}(t,\fett{q}) \, \dd q^i \dd q^j \,,
\ee
where $t$ is the proper time of the CDM particles and $\fett{q}$ are
comoving/Lagrangian coordinates, constant for each
pressureless and irrotational CDM fluid element. The gradient expansion approximates the spatial metric with a series containing an increasing number of spatial gradients expressed in powers of the initial Riemann 3-tensor and its spatial derivatives.\footnote{The Weyl tensor vanishes in three dimensions so the Riemann
tensor is fully described in terms of the Ricci tensor.}
Assuming standard inflationary initial conditions, the initial seed metric is
\be
  k_{ij} =  \delta_{ij} \left[ 1+ \frac{10}{3} \Phi(\fett{q}) \right] \,,
\ee
where $\Phi(\fett{q})$ is the primordial Newtonian potential (in our case a Gaussian field),
given at the initial time $t_0$. Using the formalism of Ref.~\cite{Rampf:2012pu}  we then obtain up to four spatial gradients\footnote{The same metric up to a transverse traceless tensor and without the decaying modes can also be found in Eq.\,(3.42) of \cite{Russ:1995eu}.}
\begin{align}
  &\gamma_{ij} (t,\fett{q}) = a^2(t) \Bigg\{ \, \delta_{ij} \left( 1+ \frac{10}{3} \Phi \right)  \Bigg. \nonumber \\
    &\quad+ 3\,D(t) \left[ \Phi_{,ij} \left( 1-\frac{10}{3} \Phi \right)
      - 5 \Phi_{,i} \Phi_{,j} +  \frac 56 \delta_{ij}  \Phi_{,l} \Phi_{,l} \right] \nonumber \\
   &\quad  +\left( \frac{3}{2}  \right)^2 E(t)
   \Bigg[ 4\Phi_{,ll} \Phi_{,ij}   \Bigg. \nonumber
  -  \delta_{ij} \left( \Phi_{,ll} \Phi_{,mm} - \Phi_{,lm}\Phi_{,lm}  \right)   \Bigg] \nonumber \\
  &\quad + \left( \frac{3}{2} \right)^2 \left[ D^2(t) -4 E(t) \right] \,\Phi_{,li} \Phi_{,lj} + {\cal O}(\Phi^3) \Bigg. \Bigg\} \,, \label{gammaUniverseGrad}
\end{align}
where "$,i$'' denotes a differentiation with respect to Lagrangian coordinate $q_i$, summation over repeated indices is implied, and we have defined ($\Lambda \neq 0$):
\begin{align}
 \label{evoDE}
 \begin{split}
   D(t) &= \frac{20}{9} \int^t \frac{\dd t'}{a^{2}(t')} J(t')  \,, \\
   E(t)  &=  \frac{200}{81} \int^t \frac{\dd t'}{a^2(t')}   \left[ \frac{K(t')}{a^2(t')}  -  \frac{9}{10} D(t') J(t')   \right]  \,,
 \end{split}
\intertext{with}
 \label{evoJK}
 \begin{split}
  J(t) &= \left[2a(t)\right]^{-1} \int^{t}  a(t') \,\dd t' \,,  \\
  K(t) &= a(t)  \int^{t} a^{-1}(t') J^2(t') \, \dd t' \,,
 \end{split}
\end{align}
with $a(t)$ the scale factor. This result is valid for $\Lambda$CDM. 
To keep expressions as simple as possible we treat the Einstein-de Sitter (EdS) universe in the following, i.e.,
$\Omega_{\rm m} \!=\!1$, $\Omega_\Lambda\!=\!0$, and thus $a(t)= \left( t/t_0\right)^{2/3}$.
We shall generalize our findings to $\Lambda$CDM later.

The precise limits in the above time integrations mirror the chosen initial conditions which can be arbitrary.
As we shall soon see the coefficients $D$ and $E$ give the time evolution of
the displacement at linear and second order respectively, and
the velocity time coefficients are proportional to the time derivative of $D$ and $E$.
We thus need two constraints at any order, e.g., one for the initial
displacement and the other for the initial velocity (alternatively one could replace one of them with the initial acceleration field).
To get the constraint for the displacement we require at initial time
\be
 \lim_{t \rightarrow t_0} \gamma_{ij}(t,\fett{q}) = k_{ij} \,.
\ee
This can be only achieved if $D(t_0) = E(t_0)= 0$, as can be easily verified through Eq.~(\ref{gammaUniverseGrad}). As we shall see the vanishing displacement field at initial time helps us to disentangle the sources of non-Gaussianity.


The functions $J$ and $K$, given by Eq.~(\ref{evoJK}), constrain the velocity coefficients proportional to
$\dot D$ and  $\dot E$  and depend on the actual physical situation (a dot denotes a time derivative with respect to $t$). We give the general solution for $D$ and $E$ for generic initial
conditions in appendix~\ref{app:growth}. Here we report a particular compelling class since their resulting
expressions are closely related  to the "slaved'' initial conditions in
Refs.~\cite{Buchert:1992ya,Buchert:2011yu,BuchertRampf:2012,Buchert:2012mb,BuchertWiegand},
and can thus be interpreted to be of the Zel'dovich type \cite{Zeldovich:1969sb}.
Since these slaved initial conditions
require an initial velocity  which is proportional to some acceleration,
we think they model the nature of adiabatic fluctuations reasonably well.
We require for this restricted class  $D(t_0) = E(t_0) = 0$, $\dot D(t_0)= 2t_0/3$, and
$\dot E(t_0)= 2 t_0^3/21$, which leads to
\begin{align}
 \label{simpleIC}
 \begin{split}
  D(t) &= \left[ a(t) -1 \right] t_0^2   \,, \\
  E(t) &= \hskip-0.03cm\left[ -\frac 3 7 a^2(t) + a(t) - \frac 4 7 \right] t_0^4  \,,\\
 \end{split}
\end{align} 
for Eq.~(\ref{gammaUniverseGrad}).
The fastest growing modes in $D$ and $E$
do not change for any choice of initial conditions but the decaying modes do.
Our conclusions do not change for any other realistic initial conditions and hence our findings do not depend on the specific choice of the decaying modes in~(\ref{simpleIC}); we make this choice to be concrete in what follows. Our final results are also valid for $\Lambda$CDM.

\section{Newtonian coordinates and the displacement field}\label{sec:Newt}

To obtain a description in terms of the motion of particles we transform the result of the gradient expansion from
the comoving coordinates $(t,\fett{q})$ to another coordinate system $(\tau, \fett{x})$.
These two frames are connected by the coordinate transformation
\begin{align}
 \label{LtrafoGrad}
 \begin{split}
   x_i(t,\fett{q})  &= q_i + \psi_i (t,\fett{q})    \\
  \tau (t,\fett{q}) &= t+ {\cal L}(t,\fett{q}) \,,
 \end{split}
\end{align}
where $\psi_i$ and ${\cal L}$ are supposed to be small perturbations.
We shall soon identify $\fett{\psi}$ and ${\cal L}$ as the displacement field and
its velocity potential, respectively.
We thus transform the metric 
\be
  \dd s^2  = - \dd t^2 +\gamma_{ij} (t,\fett{q}) \,\dd q^i \dd q^j \,,
\ee
to a Newtonian frame
\begin{align}
  \dd s^2 = - \big[ 1 \big. &+2 \big. A(\tau,\fett{x}) \big] \dd \tau^2  \nonumber \\
   &+ a^2(\tau)  \left[ 1- 2 B(\tau,\fett{x}) \right] \delta_{ij} \dd x^i \dd x^j \,,
\end{align}
where $A$, $B$ are supposed to be small perturbations and $\gamma_{ij}$ is given in Eq.~(\ref{gammaUniverseGrad}).
Note that we have neglected the excitation of vector and tensor modes. This can be straightforwardly rectified if needed. The coordinate transformation requires
\begin{align}
  \gamma_{ij}(t,\fett{q}) &= - \frac{\partial \tau}{\partial q^i} \frac{\partial \tau}{\partial q^j}
      [1+2A(\tau,\fett{x})] \nonumber \\
  \label{tbs1Grad} &\qquad\quad  + a^2(\tau) \frac{\partial x^l}{\partial q^i} \frac{\partial x^m}{\partial q^j}
    \delta_{lm} [1-2B(\tau,\fett{x})] \,, \\
     0 &= - \frac{\partial \tau}{\partial t} \frac{\partial \tau}{\partial q^i} [1+2 A(\tau,\fett{x})] \nonumber \\
   \label{tbs2Grad} &\qquad\quad + a^2(\tau)\frac{\partial x^l}{\partial t} \frac{\partial x^m}{\partial q^i} \delta_{lm} [1-2B(\tau,\fett{x})] \,, \\
     -1 &= - \frac{\partial \tau}{\partial t} \frac{\partial \tau}{\partial t} [1+2A(\tau,\fett{x})] \nonumber \\
  \label{tbs3Grad} &\qquad\quad  + a^2(\tau) \frac{\partial x^l}{\partial t} \frac{\partial x^m}{\partial t} \delta_{lm} [1-2B(\tau,\fett{x})] \,.
\end{align}
We solve these equations for $\psi$, $\mathcal{L}$, $A$ and $B$ order by order. Formally, each small
quantity is expanded in a series, i.e.~$A = \epsilon A^{(1)} +\epsilon^2 A^{(2)}  +\ldots$,
where $\epsilon$ is supposed to be a small dimensionless parameter. The primordial potential $\Phi$ is of order $\epsilon$.

We report the results for the metric coefficients $A$ and $B$ in Appendix~\ref{app:pertNewt} since they are not needed in the following.
The coordinate transformation~(\ref{LtrafoGrad})---valid for arbitrary initial conditions but restricted to an EdS universe, is  up to second order
\begin{align}
    x_i(t,\fett{q})  = q_i &+ \frac 3 2 D \Phi_{,i}  +\left( \frac 3 2 \right)^2\!E
    \frac{\partial_{q_i}}{\fett{\nabla}_{\fett{q}}^2} \mu_2 \nonumber \\
 \label{spatial} &-5 D \partial_{q_i} \Phi^2 + \left[ 5 D + \left(\frac{v}{a}\right)^2 \right] \frac{\partial_{q_i}}{\nabq^2} C_2  \,,
\end{align}
where we can easily read off the displacement field $\fett{\psi}$ according to~(\ref{LtrafoGrad}), and
\begin{align}
   \tau(t,\fett{q}) = t &+  v \,\Phi + \left( \frac 3 2 \right)^2 a^2 \Bigg[ \frac{\dot D D}{2} \Phi_{,l}\Phi_{,l}
    + \dot E \frac{1}{\nabq^2} \mu_2 \Bigg. \Bigg.\Bigg] \nonumber \\
  &+ v  \left[ v H +\frac 3 4 a^2 \ddot D -\frac 5 3 \right] \Phi^2 \nonumber \\
 \label{time} & +v  \left[ 2 vH +3a^2 \ddot D + \frac{10}{3}\right] \frac{1}{\nabq^2}C_2 ,
\end{align}
where a dot denotes a time derivative with respect to $t$, the Hubble parameter for an EdS universe is $H =2/(3t)$,  and
we have defined
\begin{align}
 \mu_2 &\equiv \frac 1 2 ( \Phi_{,ll} \Phi_{,mm} - \Phi_{,lm} \Phi_{,lm} ) \, , \\
  v &\equiv \frac 3 2 a^2 \dot D \,,  \\
 \label{PsiConstraint} C_2 &\equiv \frac{1}{\nabq^2} \Bigg[ \frac 3 4 \Phi_{,ll} \Phi_{,mm}\!+\!\Phi_{,m} \Phi_{,llm}\!+\!\frac 1 4 \Phi_{,lm}  \Phi_{,lm}  \Bigg] \,,
\end{align}
with $1/\nabq^2$ being the inverse Laplacian.

Interestingly although not directly apparent, the kernel~(\ref{PsiConstraint}) has been also derived in Ref.~\cite{Bartolo:2010rw} where the $\alpha$ of their Eq.\,(18) is given by $\alpha \equiv - 4 \nabq^{-2} C_2$. In \cite{Bartolo:2010rw} $\alpha$ arises in the evolution equation of the second-order curvature
perturbation in the Poisson gauge.

The first line in Eq.~(\ref{spatial}) agrees with  Newtonian LPT
\cite{Buchert:1992ya,Buchert:1993xz,Bernardeau:2001qr,Rampf:2012up}; it contains the
Zel'dovich approximation with its second-order improvement (2LPT). The remnant terms are relativistic
corrections which should not bear much influence on the particle trajectories at late times. At initial time, however, they
lead to an initial displacement and thus to a  density perturbation.
Also note that the time perturbations in the first line of Eq.\,(\ref{time}) correspond to the velocity perturbations in the Newtonian approximation. Indeed, $v$ in the first line is the time coefficient of the velocity potential at leading order in LPT, and
 the bracketed term in the same line in Eq.\,(\ref{time})  leads
to $\propto g(t) \nabx^{-2}G_2$ in an Eulerian coordinate system, with
\be
 \label{G2x} G_2 = \frac 3 7 \Phi_{,ll} \Phi_{,mm} + \Phi_{,l} \Phi_{,lmm} + \frac 4 7 \Phi_{,lm} \Phi_{,lm} \,,
\ee
which is the second-order kernel for the peculiar velocity in Newtonian perturbation theory \cite{Bernardeau:2001qr}.
The remnant terms
in~(\ref{time}) are again absent in Newtonian LPT.

The precise choice of the decaying modes in $D$ and $E$ is of no great importance, reflecting only the chosen initial conditions \cite{BuchertRampf:2012}.
However, it is very important to recognize the disappearance of the Newtonian part of the
displacement field in~(\ref{spatial}) for $t\rightarrow t_0$ which occurs for \emph{any} set of decaying modes. On the other hand, the last relativistic term in~(\ref{spatial}) is in general nonvanishing for $t\rightarrow t_0$. It generates the initial displacement (we additionally take the divergence of the very equation)
\be
 \label{general}
\lim_{t \rightarrow t_0} \nabq \cdot [ \fett{x}(t,\fett{q}) -\fett{q} ]
  \equiv  v_{0}^2 C_2  \,,
\ee
with
\begin{align}
 v_{0}^2 = \left( \frac 3 2 \frac{\partial D(t)}{\partial t}\right)^2\Bigg. \Bigg|_{t=t_0}  \,.
\end{align}
For our simplified initial conditions, i.e., with the use of the growth functions~(\ref{simpleIC})
we have $v_{0}^2 \rightarrow t_0^2$. The time derivative of $D$ is proportional to a velocity potential since $D$ is the time coefficient of the displacement.
Thus, the above expression vanishes only if the initial velocity between the synchronous and Newtonian frame vanishes; this initial data however would be unphysical since we already specified a vanishing displacement field. On the contrary, for a
nonzero initial velocity the displacement kernel~(\ref{PsiConstraint}) is enhanced.

Note again that the only restriction we have made to derive Eq.\,(\ref{general}) is to
require $\gamma_{ij}(t_0,\fett{q}) = k_{ij}$ initially. In a Newtonian treatment, i.e., a coordinate transformation which relates two Euclidean metrics, the
requirement of $\gamma_{ij}^{\rm Euclidean}(t_0,\fett{q}) = k_{ij}$ would imply an exact overlapping of the Eulerian and Lagrangian frames at
initial time.
Expression\,(\ref{general}) would thus vanish.

The above result is a purely relativistic effect. It results from the
space-time mixing in the coordinate transformation~(\ref{tbs1Grad}). Note that Eq.\,(\ref{general}) is also valid for $\Lambda$CDM with the corresponding $D(t)$.

\ws

\section{Option A: Initial non-Gaussianity}\label{sec:ING}

\begin{figure}

\includegraphics[width=0.52\textwidth]{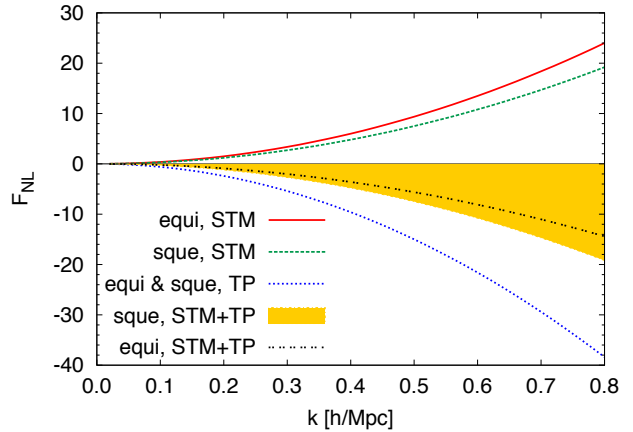}

\caption{Initial non-Gaussianity for the bispectrum~(\ref{primB}) 
for different triangle configurations with $\fett{k}_{123}\!=\!\fett{0}$.
We use the simplified initial conditions with the respective growth functions~(\ref{simpleIC}) which
yields $v_0 \rightarrow t_0$ and
$T_1 \rightarrow -1$. We choose a redshift $z=50$ such that $c t_0 \rightarrow 5.5$\,Mpc/$h$.
We plot the individual expressions in $F_{\rm NL}$ given in Eq.\,(\ref{fnlTotal}),
taking into account the non-Gaussian
amplitude induced through space-time-mixing (STM) and through the time-perturbation (TP). 
We show our results for the equilateral triangle with $k\!=\!k_1\!=\!k_2\!=\!k_3$ (equi), 
and for a squeezed triangle with $\Delta k\!=\!0.012h$/Mpc and $k\!=\!k_1\!=\!k_2$ (sque).
The red (solid) curve corresponds to the STM amplitude in the case of the equilateral triangle,
and the green (dashed) curve denotes the STM amplitude for the squeezed triangle.
 The blue (dotted) curve denotes the contribution through TP and is identical for
both squeezed and equilateral triangles. 
The black (double dotted) curve is the total amount, STM+TP,
 of non-Gaussianity for equilateral triangles. 
The total amount for squeezed triangles, STM+TP, is indicated with
the yellow (shaded) region,
since in that case the total amount can take any value.
%
Here we plot only the nonzero contributions for squeezed triangles. 
For an overview of these contributions 
see tab.\,\ref{contr}.
\label{fig:png}}
\end{figure}

The nonvanishing (relativistic) displacement field at initial time generates a density perturbation, which we shall use in the following to feed back to the Gaussian field:
\be
  \label{nonlocaldelta}  \lim_{\tau \rightarrow t_0} \delta_\Phi^{(2)} \equiv - v_{0}^2 \, C_2(\fett{q})  - \frac{v_0 t_0}{2} T_1  \nabq^2 \Phi^2 \,,
\ee
with $T_1 = - ( 1 +  10c_1 /(3 t_0^{5/3}) )$ and where $c_1$ is a constant which depends on the chosen initial conditions ($c_1 \rightarrow 0$ and so
$T_1 \rightarrow -1$ for the above mentioned simplified initial
conditions, see appendix \ref{app:growth}).
The overall minus sign is the result of mass conservation.
The last term in Eq.\,(\ref{nonlocaldelta}) is the result of using the 
Newtonian time $\tau$ and taking the limit $\tau \rightarrow t_0$ for the relativistic trajectory~(\ref{spatial}) with the use of the time
perturbation~(\ref{time}). 

 In Fourier space Eq.\,(\ref{nonlocaldelta}) becomes
\begin{align}
  \tilde{\delta}_\Phi^{(2)}(\fett{k}) &= \frac 1 4 \int \frac{\dd^3 k_1}{(2\pi)^3}
           \int \frac{\dd^3 k_2}{(2\pi)^3}
    \, (2\pi)^3 \delta_D^{(3)}(\fett{k}_{12}-\fett{k})\, \nonumber \\
   &\qquad \times F_{\rm NL}(\fett{k}_1 ,\fett{k}_2) \, \,
      \tilde\Phi(\fett{k}_1)\,  \tilde\Phi(\fett{k}_2) \,,
\end{align}
where we have defined the (already symmetrized) nonlocal kernel
\begin{align}
 \label{fnlTotal} F_{\rm NL}(\fett{k}_1,\fett{k}_2) &=  4\, v_0^2 \left[ F_{\rm NL}^{\rm STM}(\fett{k}_1,\fett{k}_2) + F_{\rm NL}^{\rm TP}(\fett{k}_1,\fett{k}_2)  \right] \,,
\end{align}
where the kernels induced through STM and through TP are respectively
\begin{align}
 F_{\rm NL}^{\rm STM} (\fett{k}_1,\fett{k}_2) &=  \left(\frac{k_1 k_2}{k_{12}}\right)^2 \Bigg[ \frac 3 4 +\frac 1 2 \frac{\fett{k}_1 \cdot
 \fett{k}_2}{k_1 k_2} \left( \frac{k_1}{k_2} +\frac{k_2}{k_1} \right)  \Bigg. \nonumber \\
 &\quad\hspace{2.082cm}
   + \frac 1 4 \frac{\left( \fett{k}_1 \cdot \fett{k}_2 \right)^2}{k_1^2 k_2^2} \Bigg. \Bigg]  \,, \\
 F_{\rm NL}^{\rm TP}(\fett{k}_1,\fett{k}_2) &= \frac{t_0}{2v_0} T_1  k_{12}^2 \,.
\end{align}
We use the short-hand notation $\fett{k}_{12}=\fett{k}_1+\fett{k}_2$ and $k_{12} =|\fett{k}_{12}|$.
With $\tilde\delta_\Phi^{(1)}=2\Phi$ (see the following section) we have
\begin{align}
 \tilde\delta_\Phi = \tilde\delta_\Phi^{(1)} +\tilde\delta_\Phi^{(2)} \,,
\end{align}
and  we recognize that
the bispectrum induced through STM and TP is
\begin{align}
  B_\Phi ({k}_1,{k}_2,{k}_3) &= 2 F_{\rm NL}(\fett{k}_1,\fett{k}_2)\, P_\Phi(k_1) P_\Phi(k_2) \nonumber \\
  \label{primB} &\qquad  + {\rm two~ permutations  } \,,
\end{align}
at leading order, with the power spectrum
\begin{align}
 \left\langle \tilde\Phi(\fett{k}_1) \tilde\Phi(\fett{k}_2) \right\rangle_c
  &\equiv (2\pi)^3 \delta_D^{(3)} (\fett{k}_{12}) \,P_\Phi(k_1) \,,
\end{align}
and similarly for the bispectrum.
In the following we compare $F_{\rm NL}$ from Eq.\,(\ref{primB}) 
with the $f_{\rm NL}^{\rm loc}$ parameter from  
the primordial non-Gaussianity of the local type. 
In Fig.\,\ref{fig:png} we plot the dimensionless quantity $F_{\rm NL}$. 
  We have used the simplified set of initial conditions, with growth functions
given in~(\ref{simpleIC}).
We have specified the initial time $t_0$ at a redshift $z$ according to
\be
  t_0(z) \simeq \frac{2}{3 H_{\rm today}  (1+z)^{3/2}}  \,,
\ee
for an EdS universe, where $H_{\rm today} = 100$\,km s$^{-1}h$\,Mpc$^{-1}$, and $h$ is the dimensionless Hubble parameter of order 0.7. We have fixed $z=50$ in the figure\footnote{At $z=158$ we have $c t_0 \rightarrow 1$\,Mpc/$h$.}
but any other starting point is appropriate and just leads to a rescaling of the $y$ axis with the amplitude of $F_{\rm NL}$ getting smaller with larger $z$.
In this figure we show the results
for equilateral and squeezed triangles ($\Delta k = 0.012h/$Mpc). We have separated the non-Gaussian amplitude with respect to its origin, i.e., the amplitude from
STM and from TP. The yellow (shaded) region denotes
the ``sum'' of both amplitudes for squeezed triangle configurations;
note that for squeezed triangles the amplitudes STM and TP are not additive as it is the case 
for equilateral triangles (which is denoted by the black [double dotted] curve). We
show the respective contributions to the overall non-Gaussian amplitude, Eq.\,(\ref{fnlTotal}), in the case of squeezed triangles in Table \ref{contr}.

\begin{table}[b]
\caption{Contributions to the non-Gaussian amplitude in
Eq.\,(\ref{fnlTotal}) for squeezed triangles with $ \fett{k}_1 + \fett{k}_2 +\fett{\Delta k} = \fett{0}$, and small $\Delta k$. The respective nonzero values are shown in Fig.\,\ref{fig:png}.}\label{contr}
 \begin{tabular*}{\columnwidth}{l  c  c }
 \hline \hline     & Space-time mixing (STM) & \quad Time pert. (TP) \\ \hline
   $F_{\rm NL}(\fett{k}_1,\fett{k}_2)$  & $\neq 0$ & $\rightarrow 0$ \\  
   $F_{\rm NL}(\fett{k}_1,\fett{\Delta k})$  & $\rightarrow 0$ & $\neq 0$ \\ 
   $F_{\rm NL}(\fett{k}_2,\fett{\Delta k})$  & $\rightarrow 0$ & $\neq 0$ \\
\hline \hline
 \end{tabular*}
\end{table}

Equation~(\ref{primB}) does only include the density perturbation induced through STM and TP [cf.~Eq.\,(\ref{nonlocaldelta})].
It is important to keep in mind that the initial acceleration between the frames (i.e., an initial acceleration of the particle) will induce a
density perturbation as well---due to Einstein's equivalence principle. The appearance of an initial acceleration obviously depends on the chosen initial conditions.

A precise scheme to implement this non-Gaussian component into
an $N$-body simulation can be found in Refs.~\cite{Wagner:2010me,Wagner:2011wx}. Then, the displacement and the velocity of the particles is just with respect to the Newtonian trajectory, i.e., it is given by  Newtonian 2LPT  \cite{Scoccimarro:1997gr}.

\section{Option B: relativistic trajectory}\label{sec:opB}

Here we include the relativistic corrections in the so-called active approach. We
call it active since not the Gaussian field is modified  as in in the passive approach  but the initial displacement of the particles.

The standard procedure to generate initial conditions in $N$-body simulation is to use
Newtonian 2LPT \cite{Scoccimarro:1997gr,Crocce:2006ve}. In there, particles are usually displaced
according to the fastest growing mode solutions of the second order displacement field together with its
 velocity field. Whilst incorporating relativistic
corrections, however, one should include decaying modes as well, since
they have roughly the same signature as the relativistic corrections.
Additionally and crucially,
by neglecting the decaying modes one loses the control to adjust the initial velocities of the particles. As a consequence
the magnitude of the initial velocity is rather accidental if decaying modes are neglected (since the initial velocity is then given by time differentiating of the fastest growing mode only).
The resulting statistics of the matter fields are accidental as well, so artificial 
transients could flaw the (initial)
gravitational evolution.
In Appendix
\ref{app:growth} we report the most general solution to adjust any initial data, depending on the actual
physical situation.
Here we describe how to set up generic initial data with the inclusion of (to be fixed) decaying modes together
with the relativistic corrections. As before we require $D(t_0) = E(t_0) =0$.
Note that this requirement does not
harm the initial density configuration as long as the velocity and the acceleration field is generated according to the full displacement field.

Suppose we have a simple rectangular grid in our $N$-body setting, thus without any spatial deformations.
We adjust a universal clock which
may be identified with an observers clock. We identify the universal clock to be $\tau$, given
in Eq.\,(\ref{time}). Then, the particles position~(\ref{spatial}) is given by the initial grid position
$\fett{q}_{(i,j,k)} \equiv \fett{q}$ plus the displacement at initial time (we suppress the $i,j,k$ label for the
specific grid points):
\begin{align}
 \lim_{\tau \rightarrow t_0} x_i(\tau,\fett{q})  = q_i &- \frac{v_0 t_0}{2}\!\left(\!1 +\frac{10}{3} \frac{c_1}{t_0^{5/3}} \right) \partial_i \Phi^2  
+ v_0^2 \frac{\partial_{i}}{\nabq^2} C_2 ,
\end{align}
where $c_1$ is a constant which depends on the chosen initial conditions (see Appendix \ref{app:growth}).
This is the initial displacement of the particles from its original unperturbed grid positions.
Note that at initial time the above can be either evaluated at the vicinity of the 
Eulerian or Lagrangian spatial coordinate. This is however not true for the Eulerian and Lagrangian time.
The velocity field at initial time is thus
\begin{align}
  {u}_{i}(t_0) &\equiv \lim_{\tau\rightarrow t_0}\frac{\dd {x_i}(\tau,\fett{q})}{\dd \tau} \nonumber \\
    &= v_0 \Phi_{,i}
    +\left( \frac 3 2 \right)^2  E' \frac{\partial_i}{\nabq^2} \mu_2  \nonumber \\
  &\qquad- \left[  \frac{17}{3}v_0+t_0 - \frac{20}{9} \frac{c_1}{t_0^{2/3}}  \right] \partial_i \frac{\Phi^2}{2} \nonumber \\
  &\qquad +\frac{4v_0}{3}  \left[  2 +\frac{v_0}{t_0} -10 c_1 t_0^{-5/3} \right] \frac{\partial_i}{\nabq^2}C_2    \,,
\end{align}
where a prime denotes a differentiation with respect to Eulerian/Newtonian time $\tau$.

Since we require a vanishing Newtonian displacement field at initial time we should
specify the initial acceleration field for the sake to get a physical set of initial data.
The obvious way to do so in an $N$-body simulation is simply to require an initial acceleration
component for each CDM particle. Instead of that one can take use of Einstein's equivalence
principle and translate the acceleration component into a density perturbation, i.e., allow different particle masses at different grid points. We report the required density perturbation in the following section.

It is then straightforward to include the above into $N$-body simulations.
The general procedure how to obtain the displacement with its velocity on a grid
can be found in Appendix D2 in Ref.~\cite{Scoccimarro:1997gr}.

\section{The density contrast in the Newtonian frame}\label{sec:deltaN}

For the sake of completeness we also derive the density contrast in the new coordinate system.
In the following it is convenient to restrict to the fastest growing modes only, since then we have a clean identification with
the density contrast from the (Newtonian) literature.
The spatial transformation of the fastest growing mode is
\begin{align}
   \label{SpatialPlus}  x_+^i(t,\fett{q})  &= q^i +\frac 3 2 a(t) t_0^2 \partial^{i}\Phi(t,\fett{q})
   + \partial^i \psi_+^{(2)} (t,\fett{q})  \,,
\end{align}
{with the scalar}
\begin{align}
  \label{F2plus}  \psi_+^{(2)}  &=  -\left(\frac 3 2 \right)^2 \frac 3 7 a^2 t_0^4
   \frac{1}{\nabq^2} \mu_2 
-5 a t_0^2 \Phi^2 +6 a t_0^2  \frac{1}{\nabq^2} C_2 \,,
\end{align}
and the temporal transformation is
\begin{align}
 \label{TimePlus}
  \tau_+ (t,\fett{q}) &= t+ t\,  \Phi (t,\fett{q}) + t\, L_+^{(2)}(t,\fett{q}) \,,
\intertext{with}
 \label{L2plus}  L_+^{(2)}(t,\fett{q}) &= \frac 3 4 a t_0^2 \Phi_{,l} \Phi_{,l}
-  \frac 9 7  a t_0^2 \frac{1}{\nabq^2} \mu_2  - \frac 7 6  \Phi^2   + 4  \frac{1}{\nabq^2} C_2 \,.
\end{align}
Note that the dependences in expressions~(\ref{F2plus}) and~(\ref{L2plus}) are either $(t,\fett{q})$ or
$(\tau,\fett{x})$ at second order.
The energy density $\rho$ for an EdS universe written in the synchronous gauge is \cite{Rampf:2012pu}
\be \label{rhoSync}
  \rho(t,\fett{q}) \equiv  \overline \rho (t) \left[ 1+ \delta(t,\fett{q}) \right] =
   \frac{3H_0^2}{8\pi G}
  \frac{\left[ 1+\frac{10}{3}\Phi(\fett{q}) \right]^{3/2}}{\sqrt{\det \left[\gamma_{ij}(t,\fett{q}) \right]}} \,,
\ee
where $\overline \rho$ is the mean density. We thus transform the above according to~(\ref{SpatialPlus})
and~(\ref{TimePlus}).
Then, we obtain the density contrast in the Newtonian coordinate system:
\begin{align}
 \label{deltaGRfull}  \frac{\delta\rho(\tau,\fett{x})}{\overline \rho}
    &=  \delta^{(1)}(\tau,\fett{x}) + \delta^{(2)}(\tau,\fett{x}) \,,
\end{align}
{where}
\begin{align}
 \label{deltaGR1} \delta^{(1)}(\tau,\fett{x}) &= \delta_{\rm N}^{(1)} +2 \Phi \,, \\
  \delta^{(2)}(\tau,\fett{x}) &=  \delta_{\rm N}^{(2)} +\frac{15}{4} at_0^2 \Phi_{|l}\Phi_{|l}
   + 3at_0^2 \Phi \Phi_{|ll} \nonumber \\
 \label{deltaGR2} &\qquad \hspace{1.4cm}-3at_0^2 \frac{1}{\nabx^2} G_2  + 8 \frac{1}{\nabx^2} C_2  \,,
\end{align}
$G_2$ is given in equation~(\ref{G2x}), and the Newtonian part of the densities at first and second-order are respectively
\be
 \delta_{\rm N}^{(1)}(\tau,\fett{x}) \equiv  -\frac 3 2 at_0^2 \Phi_{|ll} \,,  \qquad
\delta_{\rm N}^{(2)}(\tau,\fett{x})  \equiv  \left( \frac 3 2 \right)^2 a^2 t_0^4 F_2 \,,
\ee
with
\be
  F_2 = \frac 5 7 \Phi_{|ll} \Phi_{|mm} + \Phi_{|l} \Phi_{|lmm} +\frac 2 7 \Phi_{|lm} \Phi_{|lm} \,.
\ee
All dependences are with respect to  $(\tau,\fett{x})$, a vertical slash ${}_{|i}$ denotes a partial derivative
with respect to Eulerian coordinate $x_i$, and
we have neglected terms $\propto \Phi^2$ which are not enhanced
by spatial gradients.
It is also interesting to note the following relation:
\be
 \frac 3 2 \frac{1}{\nabx^2} G_2 = \frac 3 2 C_2
  - \left( \frac 3 2 \right)^2 \frac 3 7 \frac{1}{\nabx^2} \mu_2 \,.
\ee
It links the second-order velocity perturbation to the second-order displacement field---through $C_2$.
Thus, a term $\propto a t_0^2 C_2$ also arises in the density contrast (\ref{deltaGR2}). Its prefactor differs though in comparison with Eq.\,(\ref{nonlocaldelta}) since we neglect decaying modes in this section but also since it is derived in a different gauge.

Our result~(\ref{deltaGRfull}) seems to be in agreement with the second-order density contrast in the Poisson gauge in Ref.~\cite{Bartolo:2010rw}, i.e., we
obtain the same spatial functions after a couple of manipulations. However, the prefactors seem to disagree.
 We leave the full analytic comparison of the
second-order $\delta$ in Eq.\,(29) of Ref.~\cite{Bartolo:2010rw} for a future project.


\section{Which density is measured in Newtonian $N$-body simulations?}\label{sec:which}

In Secs.~\ref{sec:ING} and~\ref{sec:opB} we have formulated two ways how to include the relativistic corrections.
Both could be incorporated  whilst setting up the initial conditions in $N$-body simulations.
In the last section we have calculated the density contrast in the Newtonian gauge, so one might wonder which density
will be measured in $N$-body simulations. Since we assume that the onwarding gravitational evolution in such simulations is
still performed on a rectangular grid, the density contrast will be the same as in the Newtonian approximation.
If the simulation could entirely satisfy general relativity we would measure the density contrast in the Newtonian gauge.
This is so since space-time is deformed in general relativity and so is its 3-volume. Thus, to measure
the (mass) density in the Newtonian gauge, Eq.\,(\ref{deltaGRfull}),
we have to deform the volume $V_{\cal G}$ of the grid cells in the $N$-body simulation according to
\be
  V_{\cal G}(t) =  \int_{\cal G} J \, \dd^3 q \,,
\ee
with the peculiar Jacobian \cite{BuchertWiegand,Buchert:2011sx}
\be
   J := \sqrt{\det[g_{ij}] / \det[k_{ij}]} \,, \qquad g_{ij} := k_{ab} \,x_{\ws ,i}^{a} \,x_{\ws ,j}^{b} \,,
\ee
where the second-order deformation tensor $x_{i,j}$ is given by partial differentiation of Eq.\,(\ref{spatial}).
The local density in the Newtonian gauge is then just the ratio of massive particles in the grid cell to the deformed volume $V_{\cal G}$.


\section{Conclusions}\label{sec:con}

We obtained the relativistic Lagrangian displacement field together with its velocity potential from a general
relativistic gradient expansion for an Einstein-de Sitter universe.
By requiring that the metric reduces to a given initial condition in the synchronous comoving/Lagrangian frame we found a nonzero initial displacement field in the Newtonian/Eulerian frame which is absent in a purely Newtonian description.
This initial displacement depends on the initial velocity potential of the particle, at second order in the primordial potential.
We gave an example of initial conditions which are closely related to the so-called "slaved'' ones and are thus of the Zel'dovich type.
By including the decaying modes in the growth functions, we report the most general solution for the displacement and velocity in Appendix~\ref{app:growth}.

The findings allow us to study the relevance of the velocity (of the fluid particle) between the synchronous and Newtonian coordinate system. We find that an initial velocity between the frames generates a nonlocal density perturbation.
This is a purely relativistic effect which originates from two distinct sources, (1) space-time mixing in the coordinate
transformation~(\ref{tbs1Grad}), and (2) from the fact that the particles' proper time is coordinate dependent. The net effect is not a result of the gravitational nonlinear evolution,
and it is even apparent if the initial acceleration is zero. (Nonzero accelerations generate density
perturbations even in the Newtonian treatment.)
Equation~(\ref{primB}) shows that the coordinate transformation induces a small velocity dependent amount of non-Gaussianity whose scale dependence is depicted
in Fig.\,\ref{fig:png}.
Our result could be of importance in situations where the velocities of the traced
objects are relatively high, since the $f_{\rm NL}$ amplitude depends on the particle velocity potential squared.

We considered the generation of initial conditions in $N$-body simulations which we described in Sec.~\ref{sec:ING}: Instead of
using the relativistic trajectory~(\ref{spatial}) one can treat the
relativistic corrections in terms of an initial non-Gaussianity component, which acts as a correction to
the (Lagrangian) potential and thus as a correction to the initial Gaussian field. We therefore adjust the
cosmological potential with respect to the initial nonlocal density perturbation but then let the
particles evolve according to the Newtonian trajectory. This establishes a simple quasirelativistic
$N$-body simulation, and we call it the passive approach.
In Sec.~\ref{sec:opB} we described the active approach---the alternative way to incorporate the relativistic corrections in $N$-body simulations.
In the active approach, the initial Gaussian field is unaltered but the particle trajectories include relativistic corrections.
As an important note, we assume that the resulting initial
statistics in both approaches are equivalent, leave however the cumbersome proof for future work. If the resulting statistics agree,
the same Lagrangian method could be used to generate non-Gaussian initial conditions for purely Newtonian $N$-body simulations.

Finally, in section \ref{sec:which} we showed how to relate the density contrast from $N$-body simulations to the one in the Newtonian gauge. Essentially, space-time is deformed and so is the 3-volume of the grid cells. Thus, the density in the Newtonian gauge is obtained by relating it to the appropriate physical volume.

\acknowledgments

A part of this work contributed to the dissertation of 
C.R.~at RWTH Aachen University \cite{Rampf:2013thesis}.
C.R.~would like to thank Y.\,Y.\,Y.\,Wong for valuable discussions, and A.\,Oelmann for comments on the manuscript. G.R.~is supported by the Gottfried Wilhelm Leibniz programme of the Deutsche Forschungsgemeinschaft (DFG).




\appendix

\section{General growth functions}\label{app:growth}

The general solutions for the displacement coefficients~(\ref{evoDE}) in an EdS universe are
\begin{align}
 D(t) &= \left[ t^{2/3} - \frac{20}{9} \frac{c_1}{t} + \frac{20}{9} c_2 \right] t_0^{4/3} \,,\\
 E(t) &=  \Bigg[ -\frac{3}{7} t^{4/3}  -\frac{20}{9} c_2 t^{2/3}
   -\frac{40}{9} \frac{c_1}{t^{1/3}}  -\frac{200}{81} \frac{c_3}{t}  \Bigg. \nonumber \\
  &\qquad
   +\frac{400}{81} \frac{ c_1 c_2}{t} -\frac{100}{81}  \frac{c_1^2}{t^2}
   +\frac{200}{81} c_4 \Bigg. \Bigg] t_0^{8/3} \,,
\end{align}
where $c_1$ -- $c_4$ are constants. To fix the constants one may choose specific initial data for
$D(t_0)$ and $E(t_0)$  together with its first time derivatives. Again, this conforms to specify the
initial displacement and the initial velocity.

Generally it could be also appropriate to fix the initial peculiar gravitational
field instead of the initial velocity,
since the former sources a density perturbation and thus could flaw our main argument in this paper (i.e., that the initial non-Gaussianity is sourced only through initial velocity perturbations). Indeed, defining the
time coefficient of the initial peculiar gravitational field as
\be
   g_{\rm pec} (t_0) = \frac{4}{3t_0} \dot D(t_0) + \ddot D(t_0) \,,
\ee
up to first order (similarly for the higher order coefficients),
we can construct displacements which
are entirely without initial accelerations. We call this initial data "inertial'', and the resulting displacement coefficients are then
\begin{align}
 \label{Dinert} D_{\rm inert}(t) &= \left[ a - a^{-3/2} \right] t_0^2  \,, \\
 \label{Einert} E_{\rm inert}(t) &= \left[ -\frac 3 7 a^2 -2 a^{-1/2} +\frac{75}{28} a^{-3/2} -\frac 1 4 a^{-3}  \right] t_0^4 \,.
\end{align}
As can be proven by direct verification, the above modes do not change our conclusions: the coordinate
transformation yields to a density perturbation which is only sourced by a velocity perturbation.

\section{Perturbations in the Newtonian metric}\label{app:pertNewt}
Here we summarize our findings for the metric
\begin{align}
  \dd s^2 &= - \left[ 1+2 A(\tau,\fett{x}) \right] \dd \tau^2  \nonumber \\
   &\qquad \quad+ a^2(\tau)  \left[ 1- 2 B(\tau,\fett{x}) \right] \delta_{ij} \dd x^i \dd x^j \,.
\end{align}
To avoid cumbersome expressions we restrict to an EdS universe and take only the fastest growing modes into
account.
Up to second order in $\Phi$ the metric coefficients are
\begin{align}
 &A(\tau,\fett{x}) \simeq {\phi}_{\rm N} - {\phi}_{\rm R}  \,,\\
\intertext{and}
&B(\tau,\fett{x}) \simeq {\phi}_{\rm N}   + \frac 2 3 {\phi}_{\rm R}  \,,
\end{align}
with
\begin{align}
 {\phi}_{\rm N} &=-\Phi +\frac 3 4 a(\tau) t_0^2 \Phi_{|l}  \Phi_{|l} 
+ \frac{15}{7} a(\tau) t_0^2 \frac{1}{\fett{\nabla}_{\fett{x}}^2} \overline\mu_2 \,, \\
 {\phi}_{\rm R} &=  \frac{4}{\nabx^2 \nabx^2} \left[ \frac 3 4 \Phi_{|ll} \Phi_{|mm}
   + \Phi_{|l} \Phi_{|lmm} + \frac 1 4 \Phi_{|lm} \Phi_{|lm} \right]  \,, \\
 \overline\mu_2  &=  \frac 1 2 \left( \Phi_{|ll} \Phi_{|mm} - \Phi_{|lm} \Phi_{|lm} \right) \,.
\end{align}
A vertical slash "$|_{i}$'' denotes a differentiation w.r.t.~Eulerian coordinate $x_i$.
Note that we have $\Phi \equiv \Phi(t_0,\fett{x})$ for this appendix.
The term ${\phi}_{\rm N}$ can be obtained from a purely Newtonian treatment
\cite{Stewart:1994wq,Rampf:2013dxa,Rampf:2013thesis}, and it is just the cosmological potential up to second order.
Indeed, multiplying the second term on the rhs with $\nabx^2/ \nabx^2$ we find
\be
   {\phi}_{\rm N} =-\Phi + \frac 3 2 a(\tau) t_0^2 \frac{1}{\nabx^2} F_2(t_0,\fett{x}) \,,
\ee
with the well-known second-order kernel
\be
   F_2(t_0,\fett{x}) = \frac 5 7 \Phi_{|ll} \Phi_{|mm}
   + \Phi_{|l} \Phi_{|lmm} + \frac 2 7 \Phi_{|lm} \Phi_{|lm}  \,.
\ee


\begin{thebibliography}{99}
\bibitem{Mukhanov:1990me} V.\,F.\,Mukhanov, H.\,A.\,Feldman, and
                     R.\,H.\,Brandenberg-er,
     Phys.\,Rep. \textbf{215}, 203 (1992).

\bibitem{Ma:1995ey} C.-P.\,Ma,  and E.\,Bertschinger, {Astrophys.\,J.} \textbf{455},
 7 (1995).

\bibitem{Maldacena:2002vr} J.\,M.\,Maldacena, {J.\,High Energy Phys.} \textbf{05}, (2003) 013,
 [astro-ph/0210603].

\bibitem{Acquaviva:2002ud} V.\,Acquaviva, N.\,Bartolo, S.\,Matarrese, and A.\,Riotto,
  {Nucl.\,Phys.} \textbf{B667}, 119 (2003), [astro-ph/0209156].

\bibitem{Bartolo:2004if} N.\,Bartolo,  E.\,Komatsu, S.\,Matarrese, and A.\,Riotto, {Phys.\,Rep.}
 \textbf{402}, 103 (2004), [astro-ph/0406398].

\bibitem{Bartolo:2006fj} N.\,Bartolo, S.\,Matarrese,  and A.\,Riotto, {JCAP} \textbf{0701}, 019 (2007), [astro-ph/0610110].

\bibitem{Beneke:2010eg} M.\,Beneke and C.\,Fidler, {Phys.\,Rev.} \textbf{D82}, 063509 (2010), [arXiv:1003.1834].

\bibitem{Pettinari2013} G.\,W.\,Pettinari, C.\,Fidler,  R.\,Crittenden, K.\,Koyama,  and D.\,Wands, (2013), [arXiv:1302.0832].

\bibitem{Hu:1993tc} W.\,Hu, D.\,Scott, and J.\,Silk,  {Phys.\,Rev.} \textbf{D49}, 648 (1994), [astro-ph/9305038].

\bibitem{DeTroia:2003tq} G.\,De Troia,  P.\,A.\,R.\,Ade,  
J.\,J.\,Bock, J.\,R.\,Bond,   A.\,Bos-caleri, \emph{et al.}, {Mon.\,Not.\,R.\,Astron.\,Soc.}  \textbf{343}, 284 (2003),
      [astro-ph/0301294].

\bibitem{Bernardeau:2001qr}
F.\,Bernardeau, S.\,Colombi, E.\,Gaztanaga, and R.\,Scocci-marro, {Phys.\,Rept.}  \textbf{367}, (2008)
 [astro-ph/0112551].

\bibitem{Matarrese:1995sb} S.\,Matarrese and D.\,Terranova,  {Mon.\,Not.\,R.\,Astron.\,Soc.} \textbf{283}, 400 (1996), [astro-ph/9511093].

\bibitem{Rampf:2012pu} C.\,Rampf and G.\,Rigopoulos,
 {Mon.\,Not.\,R.\,Astron.\,Proc.} \textbf{430}, L54
 (2013), [arXiv:1210.5446].

\bibitem{Lifshitz:1963ps}
   E.\,M.\,Lifshitz and I.\,M.\,Khalatnikov, 
 {Adv.\,Phys.}  \textbf{12}, 185 (1963).

\bibitem{Tomita:1975kj} K.\,Tomita, 
{Prog.\,Theor.\,Phys.} \textbf{54}, 730  (1975).

\bibitem{Tanaka:2006zp} Y.\,Tanaka and M.\,Sasaki, 
  {Prog.\,Theor.\,Phys.}  \textbf{117}, 633 (2007).

\bibitem{Stewart:1994wq} J.\,M.\,Stewart, D.\,S.\,Salopek,  and K.\,M.\,Croudace,
    {Mon. Not. R. Astron. Soc.} \textbf{271}, 1005 (1994),  [astro-ph/9403053].

\bibitem{Enqvist:2011aa} K.\,Enqvist,  S.\,Hotchkiss, and G.\,Rigopoulos, {JCAP} \textbf{1203},
   026 (2012), [arXiv:1112.2995].

\bibitem{Rigopoulos:2012xj}  G.\,Rigopoulos and W.\,Valkenburg, {Phys.\,Rev.} \textbf{D86}, 043523  (2012), [arXiv:1203.2796].

\bibitem{Bartolo:2010rw} N.\,Bartolo, S.\,Matarrese, O.\,Pantano, and A. Riotto,
 {Class.\,Quant.\,Grav.} \textbf{27}, 124009 (2010), [arXiv:1002.3759].

\bibitem{Matarrese:1994wa} S.\,Matarrese,  O.\,Pantano, and D.\,Saez, {Mon. Not. R. Astron. Soc.} \textbf{271},
513 (1994), [astro-ph/9403032].

\bibitem{Russ:1995eu} H.\,Russ, M.\,Morita,  M.\,Kasai, and G.\,Borner,
 {Phys.\,Rev.} \textbf{D53}, 6881 (1996), [astro-ph/9512071].

\bibitem{Buchert:2012mb} T.\,Buchert and M.\,Ostermann, {Phys.\,Rev.} \textbf{D86},
 023520 (2012), [arXiv:1203.6263].

\bibitem{BuchertWiegand} T.\,Buchert, C.\,Nayet, and A.\,Wiegand, 
 (2013), \\
\ [arXiv:1303.6193].

\bibitem{Buchert:1992ya}  T.\,Buchert, {Mon.\,Not.\,R.\,Astron.\,Soc.} 
\textbf{254}, 729 (1992).

\bibitem{Buchert:2011yu} T.\,Buchert, {Class.\,Quant.\,Grav.} \textbf{28}, 164007 (2011), [arXiv:1103.2016].

\bibitem{BuchertRampf:2012} C.\,Rampf and T.\,Buchert, {JCAP} \textbf{1206},  021 (2012), [arXiv:1203.4260].

\bibitem{Zeldovich:1969sb} Ya.\,B.\,Zeldovich, {Astron.\,Astrophys.} \textbf{5}, 84 (1970).

\bibitem{Buchert:1993xz} T.\,Buchert and J.\,Ehlers, {Mon.\,Not.\,R.\,Astron.\,Soc.} \textbf{264}, 375 (1993).

\bibitem{Rampf:2012up} C.\,Rampf, JCAP \textbf{1212}, 004 (2012), 
[arXiv:1205.5274].


\bibitem{Wagner:2010me} C.\,Wagner, L.\,Verde,  and L.\,Boubekeur,
    {JCAP} \textbf{1010}, 022 (2010), [arXiv:1006.5793].

\bibitem{Wagner:2011wx} C.\,Wagner and L.\,Verde,
  {JCAP} \textbf{1203}, 002 (2012), [arXiv:1102.3229].

\bibitem{Scoccimarro:1997gr} R.\,Scoccimarro, {Mon.\,Not.\,R.\,Astron.\,Soc.} \textbf{299}, 1097 (1998), [astro-ph/9711187].

\bibitem{Crocce:2006ve} M.\,Crocce, S.\,Pueblas, and R.\,Scoccimarro, {Mon. Not. R. Astron. Soc.} \textbf{373},
 369 (2006), [astro-ph/0606505].


\bibitem{Buchert:2011sx} T.\,Buchert and S.\,Rasanen,  {Ann.\,Rev.\,Nucl.\,Part.\,Sci.} \textbf{62},
 57 (2012), [arXiv:1112.5335].

\bibitem{Rampf:2013dxa} C.\,Rampf, (2013), [arXiv:1307.1725].

\bibitem{Rampf:2013thesis} C.\,Rampf, doctoral thesis, RWTH Aachen, (2013).


\end{thebibliography}
\end{document}